# The individual income distribution in Argentina in the period 2000-2009. A unique source of non stationary data.

Juan C. Ferrero[*]

*Centro Laser de Ciencias Moleculares. INFIQC and Facultad de Ciencias Químicas, Universidad Nacional de Córdoba, 5000 Córdoba, Argentina*

The economic crisis in Argentina around year 2002 provides a unique opportunity for Econophysics studies. The available data on individual income are analyzed to show that they correspond to non stationary states. However, the rather restricted size of the data survey imposes difficulties that must be overcome through a careful analysis, for a reliable use. A new method of data treatment is presented that could be helpful in theoretical studies.

**I. Introduction**

The empirical data on individual income distribution presents, for every country, two special features. One of them is the tail of the distribution, which is linear in a logarithmic graph and the other is the stability of the shape of the distribution, which changes very slowly over time and shows no appreciable differences in consecutive years. The first of these characteristics has been the subject of numerous studies[2 - 4] since first observed by Pareto, more than 100 years ago [5]. The Pareto tail of the cumulative distribution follows a power law of the type

$$Q = A \, x^{-\alpha} \qquad (1)$$

where $\alpha$ is the Pareto index, whose value is usually around 2. A notable feature of the Pareto tail that allows for a differentiation from the rest of the distribution is the change in the value of the slope with the economic situation[1]. The second indicates that the income distribution is in a quasi-stationary state, which enables model calculations based on obtaining distribution functions corresponding to steady states. This is the central assumption underlying the models developed by the Kolkata school and other researchers[6-9]. The model, based on the kinetic theory of ideal gases has been extremely successful in accounting for the main features of empirical data by introducing a saving propensity factor in the transactions between pairs of agents[10]. The model has been extensively studied and is able to reproduce the main features of the empirical distribution[11,12].

---

[*] Email: jferrero@fcq.unc.edu.ar

Since both theory and empirical data correspond to stationary situations, none of them provides information on the evolution of income and factors affecting it. This would require the availability of statistical information obtained in an economy undergoing a major perturbation in a very short period that resulted in a rapid transfer of resources between the various groups that compose it, leading to a distribution out of equilibrium and evolving towards a new steady state.

A recent case, unique from the great crises in the first half of the 20$^{th}$ century has been the economic and financial collapse that occurred in Argentina in 2002 and that developed within months. This article will examine critically the individual income data from Argentina in the years around the 2002 crisis, in order to explore the potentiality of its use in Econophysics studies.

**II. The income distribution in Argentina en the period 2000 -2009**

A succinct account on the economic crisis in Argentina, which peaked in the period December 2001 – June 2002 will serve to put the subject in perspective. It could also seem familiar to those acquainted with the present situation in Europe. Briefly, until the end of 2001 the price of local currency (the peso) was linked to the U.S. dollar in a 1:1 ratio, and there was free convertibility between the two currencies, so that all transactions were made with any of them. The rapidly deteriorating economic situation due to the overwhelming foreign debt, led to a crisis that ended with a default and the collapse of the financial system. The end of convertibility and the deterioration of the value of money quickly exceeded government expectative, reaching a ratio of 1:4, to finally stabilize at around 1:3. This resulted in a rapid, massive transfer of resources between various sectors, mainly associated with the sudden change of the local currency value. The rapid rise in unemployment and the consequent increase in poverty led the government to take palliative measures. Slowly, compared with this process, the economy evolved to a new state of equilibrium.

An analysis of the evolution of a system like this requires, from the point of view of Econophysics, empirical data of high quality and reliability.

In Argentina, through the statistical studies office, INDEC, the government collects and compiles data on population and economy through surveys on selected samples. Although statistically significant, they are necessarily limited in size. This feature establishes a difference with the empirical data from those countries where each resident must complete an annual declaration of income, which provides a very large database. One of the problems that arises from a small sample originates in the low value of the cumulative probability of the Pareto tail, which starts at typical to values below 1%, so if the database is small, this tail is not sufficiently sampled.

Individual income data, either theoretical or empirical, are usually presented as the cumulative distribution function (CDF), as in Figure 1, or the probability density function (PDF). If the amount of data is low, it seems better to use the CDF. However, in most of the curve, corresponding to the low and medium income sectors, the CDF is relatively insensitive to changes in the economic situation while the tail of the distribution may not be adequately represented. This seems to be the case in the income distribution of Argentina in years 2000 and 2001, where the change of slope that indicates the onset of the Pareto tail is not detected (Figure 1) although it is clearly observed in the post crisis years. On the other hand, the PDF is more sensitive in the region of low and middle income, but if the amount of data is low, the curve is very noisy, which aggravates the often presented problem of artificial behaviors, depending on how the data are processed.

The data for Argentina in the years studied show several interesting features that should be analyzed. In the low and medium income region the PDF is, at least, bimodal[13]. In the high-income limit in 2000 and 2001( Figure 1), the graph tends to linearity, but without showing the typical change of slope. The values of $\alpha$ for the period 2000 -2009 are presented in Figure 2. Before the peak of the crisis, the values of     are abnormally large. This could be due to the lack of data in the survey for the tail of the distribution or to a feature imposed by the on coming crisis. During the peak of the crisis, in 2002, there is a strong oscillation between extremes, which quickly tends to stabilize at around 2, indicating that the crisis has been resolved in about four years.

**III. Application to econophysics analysis**

The data in Figure 1 does not seem to present a variation consistent with the major changes in economic conditions that caused the crisis. This is a clear example of the insensitivity of the cumulative distribution function. In addition, the strong variation of     during the crisis represents a too restricted piece of data. To use this information in Econophysics, it seems therefore necessary to find a way to treat the data able to reveal the differences in income distribution when a steady state has been not reached.

An efficient method often used in experimental physics is to measure the relative values of the signals. In the case of the income distribution functions they should be calculated relatively to a year with an equilibrium distribution, taken as reference. Since the cumulative distributions corresponding to 2000 and 2001 are almost indistinguishable, one can consider that in those years there was a equilibrium state, and use these distributions as a reference. The results obtained with reference to the year 2000 are shown in Figure 3. The distribution of the last quarter of 2001 shows a significant change at low income levels, not detected in Figure 1. There is a decline in the fraction of agents in the low income region, at values lower than 100. This is not a sign of an improvement in distribution, but rather an indication that individuals in this region became unemployed and therefore with zero income. This situation worsened, as revealed by the survey data, in May 2002, which also shows an increase in the upper end of income, that corresponds to individuals who have benefited from the crisis, specifically, the local currency devaluation. By October of that year the distribution has changed with respect to to year 2000 has changed completely, showing a significant increment at both ends of the data. These relative data as well as the CDFs of Figure 1 show a shift to higher values on the income scale, which seems to be mainly due to inflation. However the curves have similar shapes, indicating that a new stationary state has been attained, with a higher level of inequity.

The sensitivity of the data to temporary changes should allow for efficient modelling studies. However it must be noted that model calculations are restricted to a constant population and a constant amount of money. Neither of them strictly holds in the case of the Argentina crisis. Although the total population of the country changed very little over this period, unemployment variations were significant and affected not only the income statistics but also the number of agents economically active. In addition, the fast recovery of the economy was based on quite favourable external markets, which resulted in a large flow of money into the country, so that the condition of constant total money should be somehow released. An additional complication is that statistical models produce equilibrium distributions that are independent of the initial state and of the path to reach it[14].

In conclusion, although the crisis in Argentina around year 2002 provides, if properly analysed, adequate economic data for an out-of-equilibrium economic state, the use of the gas kinetic model requires an extension beyond the usual constraints of a closed system.

**Acknowledgements**: The data were provided by the INDEC. This work was supported by CONICET and FONCYT (Argentina)

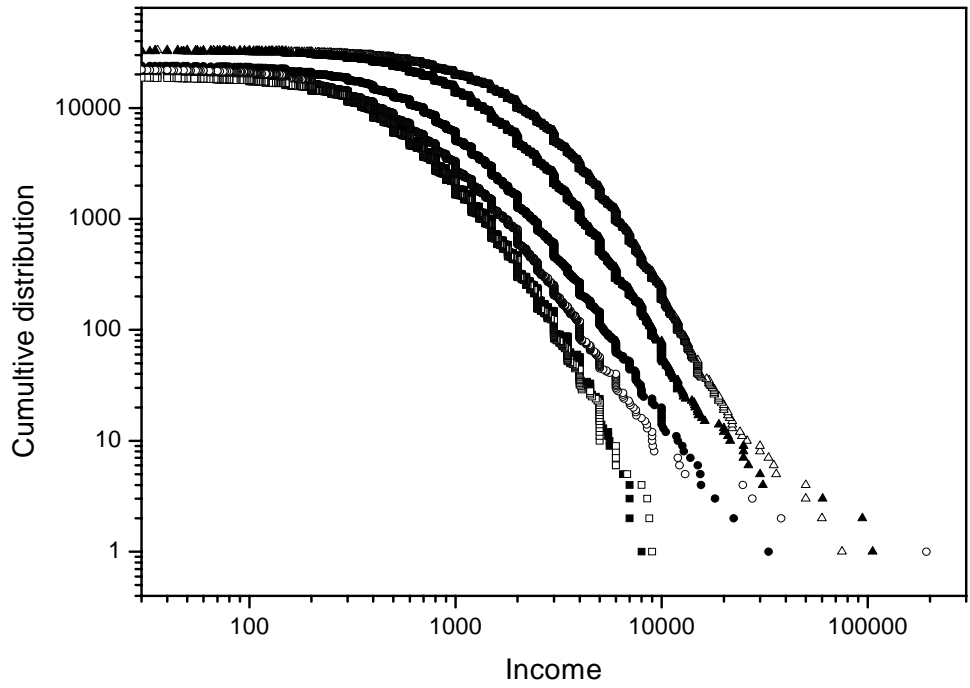

Figure 1: The non normalized cumulative distribution of income for various years:
( ■ ) 2000; ( □ ) May 2002; ( ○ ) Oct 2002; ( • ) 2003; ( ▲ ) 2005; (Δ ) 2009

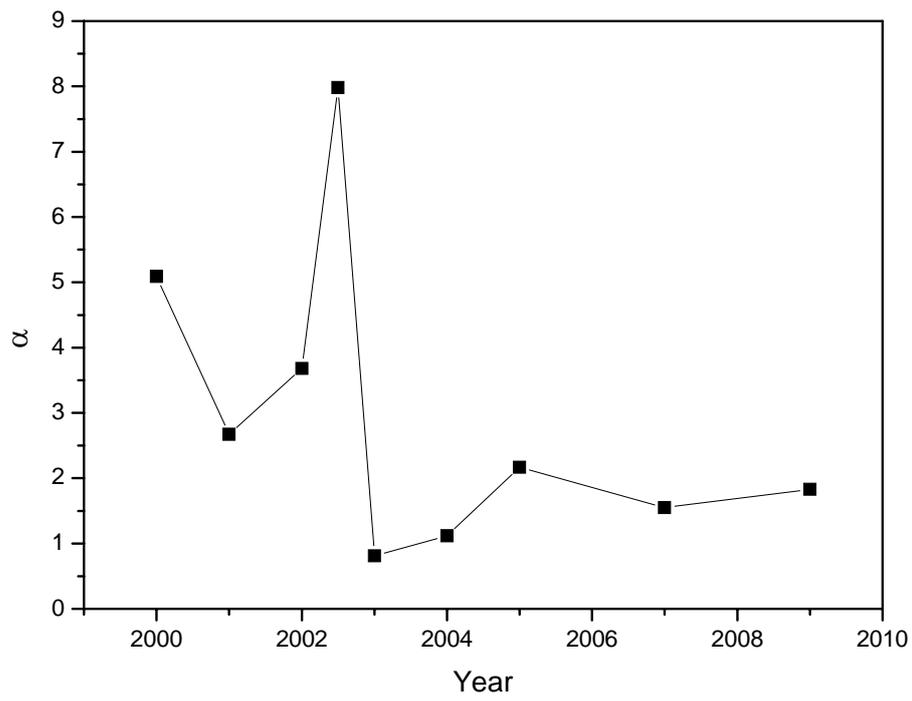

Figure 2: The Pareto exponent ( Eq. 1) as a function of year.

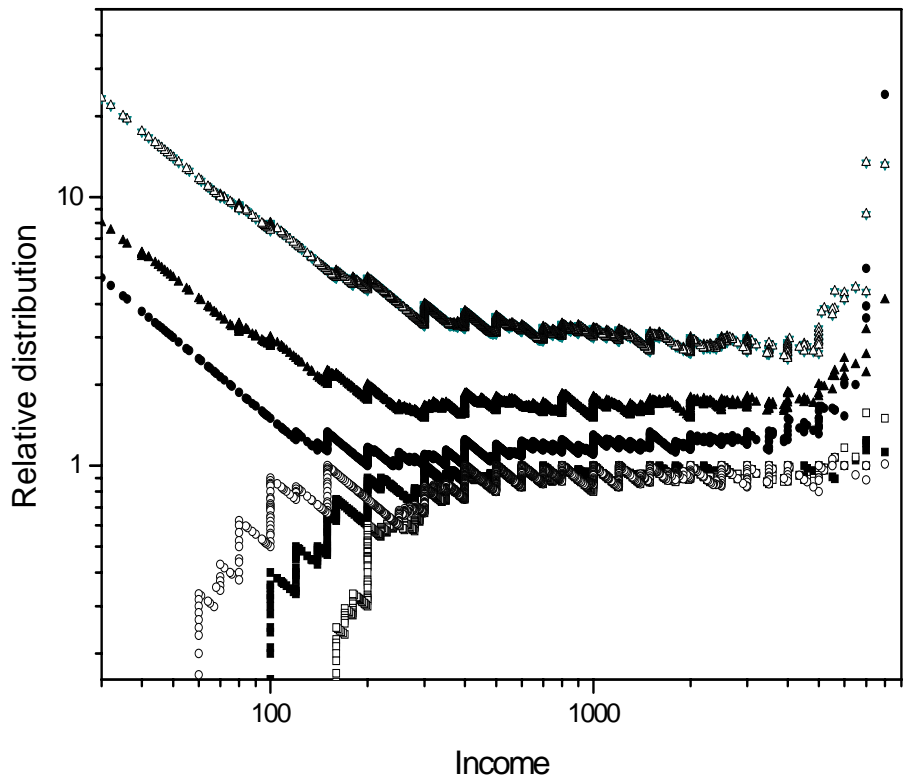

Figure 3: The cumulative distribution of income for various years relative to year 2000: The symbols are the same than in Figure 1

( ■ ) 2000; ( □ ) May 2002; ( ○ ) Oct 2002; ( • ) 2003; ( ▲ ) 2005; (Δ) 2009